# Comparative study of Mercury's perihelion advance

S.P. Pogossian,  Univ. Brest, CNRS, IRD, Ifremer, *Laboratoire d'Océanographie Physique et Spatiale* (LOPS), IUEM, 29280, Brest, France.

**Abstract**

Mercury's motion using numerical methods in the framework of a model including only the non-relativistic Newtonian gravitational interactions of the solar system: 9 planets in translation (including Pluto) around the sun has been studied. Since the true trajectory of Mercury is an open, non-planar curve, we have paid special attention to the exact definition of the advance of Mercury's perihelion. For this purpose, we have introduced the notions of an extended and a geometrical perihelion. In addition, for each orbital period, a mean ellipse was fitted to the trajectory of Mercury. I have shown that the perihelion advance of Mercury deduced from the behavior of the Laplace-Runge-Lenz vector, as well as the extended and geometrical perihelion advance depend on the fitting time interval and for intervals of the order of 1000 years converge to a value of  532.1" per century.   The behavior of the perihelia, either extended or geometrical, is strongly impacted by the influence of Jupiter. The advance of the extended perihelion depends strongly on the time step used in the calculations, while the advance of the geometrical perihelion and that deduced by the rotation of the Laplace-Runge-Lenz vector depends only slightly on it.

# 1. Introduction

The numerical determination of planetary orbits is one of the important tasks of modern astrophysics [(Seidelmann, United States Naval Observatory & Great Britain 1992), (Gurfil & Seidelmann 2016)]. In solar system calculations, the orbit of Mercury is of constant interest because it is currently accepted that the anomalous advance of Mercury's perihelion can only be explained on the basis of general relativity [ (Clemence 1947), (Rana 1987), (Stump 1988), (Morrison & Ward 1975), (Smulsky 2011)]. Moreover, the advance of Mercury's perihelion is considered as one of the tests proving Einstein's theory of general relativity. And yet, up to now initiatives have been taken to shed light on some of the limitations arising from the ambiguous definition of Mercury's perihelion advance.

The central question is thus to know what kind of physical quantity is considered as being the perihelion advance and how can this be measured [(Davies 1983), (Balogh & Giampieri 2002), (Lo, Young & Lee 2013), (Roy 2015), (Roseveare 1982), (Park et al. 2017), (Inoue 1993), (Shapiro et al. 1972), (Anderson et al. 1996)]. The various measurements of the perihelion advance are not perfect and have inevitable experimental errors which contribute to the uncertainty in determining this quantity [(Cicalò et al. 2016), (Verma et al. 2014)]. Optical observations of Mercury are subject to errors arising from the determination of the planet's barycenter from its visible disk [(Fienga 1999), (Solomon et al. 2007)]. In radar measurements, uncertainties may appear due to the very weak detection signal depending on the topography of the planet, the speed, and the relative positions of the observer and the planet. As the measurement results are analyzed using one of the existing models to determine the orbital parameters, these parameters will depend on the chosen model [(Mazarico et al. 2014), (Verma et al. 2014)]. Currently, the post-Newtonian relativistic terms are automatically included in the equations of motion for the determination of the ephemeris of Mercury, which contributes to the value of the perihelion advance [(Clemence 1948), (Poisson & Will 2014)].

In the comparisons of the Mercury perihelion advance by the different relativistic and non-relativistic models, the astronomical constants and ephemerides used in the orbital calculations based on the numerical integration of the Newtonian gravitational equations for N-bodies should be used without relativistic corrections [(Le Guyader 1993), (Arminjon 2004)]. The perihelion advance of Mercury is very small, on the order of 575" per century, and was discovered by Le Verrier in 1859 [(Le Verrier 1859)]. By processing the MESSENGER ranging data, Park et al [(Park et al. 2017)] have estimated the total precession rate of Mercury's perihelion of 575.3100±0.0015 ″/century. In their estimation, Park et al.[(Park et al. 2017) ] used the parameterized post-Newtonian formalism (PPN) formalism with fitted PPN parameters in order to extract the Mercury's precession rate from the MESSENGER ranging data.

The correction provided by general relativity is smaller and is about 42.98" per century [(Nobili & Will 1986), (Biswas & Mani 2005)] . The analysis of radar ranging data between 1966-1988 by Anderson et al[(Anderson et al. 1991)] gives a slightly different value for the excess precession of 42.94"/cy. Their difference 532.37 "/ cy = 575.31" / cy - 42.94 "/ cy is due to influences from other planets and can be explained by Newton's classical theory of gravitation [(Krızek 2017), (Loskutov 2011)]. Different values for the perihelion advance due to influences from other planets have been published by different authors, for example : 531.54"/cy [(Misner et al. 2017)], 531.5"/cy [(Clemence 1947)], 531.4"/cy [(Rydin 2011)], 531.25"/cy [(Bretagnon 1982)] , 530"/cy [(Vankov 2010)], 528.95 "/cy [(Narlikar & Rana 1985)], 528.93"/cy [(Rana 1987)], 526.7"/cy [(Le Verrier 1859)].

Quantities that are defined in an ambiguous way can introduce uncertainties that may confuse their study. It is therefore very important to determine exactly what the perihelion advance of Mercury means for theorists, and by what methods they have calculated it in order to be able to compare their values with experimental results. Ambiguities are often found in the definitions of what is called perihelion of an orbit, using the elliptical trajectory characteristic of the two-body problem. When considering the N-body problem using the extension of the parameters characterizing the two-body motion, special attention should be paid to their physical meaning. It is for these reasons that I will reconsider in this paper a certain number of concepts which are associated with the advance of Mercury's perihelion but which have an exact definition only in the framework of the two-body problem. Their possible extensions to the N-body problem and the uncertainties and approximations that might arise from such a procedure are discussed.

I will draw special attention to the relative positions of the barycenter of the solar system, the barycenter of the two-body system of the Sun and Mercury, and the positions of the Sun and the foci of the mean elliptical trajectory of Mercury (the closest focus to the Sun). The concepts of extended and geometrical perihelia are introduced in order to give a distinct account of their advance and to compare them to the value of the perihelion advance that has been determined from the evolution of the direction of the Laplace-Runge-Lenz vector (LRL) [(Stewart 2005), (Goldstein 1975), (Goldstein, Poole & Safko 2008)].

## 2. Calculations

In present calculations, only the non-relativistic Newtonian gravitational interactions of the solar system: 9 planets in translation (including Pluto) around the sun have been included. The 10-body Newtonian gravitational equations (including the sun) have been integrated over a time interval of 262144 days with a time step of about 42 minutes. It should be noted that when using this integration time interval, Mercury makes about 2980 revolutions around the sun, Venus 1167 and the Earth-Moon system 718.

The configuration of the solar system's four outer planets has a periodicity of about 178.7 years (62270 days). Jose came to this conclusion [(Jose 1965)] by examining calculations of the solar orbit around the barycenter of the solar system. The trajectory of the Sun is considerably influenced by massive and distant planets. The period defined by Jose is slightly larger than a period of Neptune around the sun allowing 2, 6 and 15 complete revolutions of Uranus, Saturn and Jupiter respectively. The perihelion advance of Mercury in a time interval equal to about four Jose cycles has been analyzed and only then the average perihelion advance per century has been calculated.

For the integration of 10-body solar system, one needs to know the values of the basic parameters of the 10-body problem: the planet/sun mass ratio, the Newtonian constant of universal gravitation, as well as the initial values of the positions and velocities of all the planets and the sun. Even if the most available ephemerides are based on experimentally measured data, they include corrections based on general relativity, since the values of all the parameters depend on the model-dependent optimization procedures. Thus, Newtonian ephemeris data without relativistic corrections are required for calculations. Le Guyader used an optimization program to subtract the relativistic corrections and the influence of the moon which correspond to the particular values of the gravitational constants (GM) taken from the DE200/LE200 ephemeris [(Le Guyader 1993)]. In order to obtain ephemerides without relativistic corrections, Arminjon et al [(Arminjon 2004)] went further by simultaneously fitting the planets' gravitational constants to their initial positions and velocities. For pure Newtonian calculations, Roy (Roy 2014) used the initial positions and velocities of the planets without the relativistic corrections as calculated by Le Guyader. But for gravitational constants Roy (Roy 2014) has used the IAU 1976 values from Taff's book [(Taff 1985)]. In the choice of the initial values for the numerical calculations, it is necessary to have coherence between the initial values of the orbital state vectors (three-dimensional vector for the position and for the velocity) and the gravitational constants. In this study the initial positions and velocities optimized by Le Guyader have been used, as far as the gravitational constants and the astronomical unit are concerned, their values as reported in the ephemeris DE200/LE200 [(Le Guyader 1993), (Standish 1990) ] and which were used by Le Guyader during the subtraction of the relativistic corrections have been used.

A MATLAB ODE113 solver was used with RelTol =3 ×10$^{-14}$ and AbsTol= 10$^{-16}$, values that were slightly different from those recommended by Arminjon [(Arminjon 2004)] but which provide a better accuracy for the initial data first used by Le Guyader [(Le Guyader 1993)]. The coordinates and initial velocities of the sun and the 9 planets (with Pluto) at the date of the Julian ephemeris JJ = 2451600.5 taken from the reference [(Le Guyader 1993)] were used. The integration results were analyzed in an invariant coordinate system related to the initial value of the barycentric linear momentum vector $\vec{P} = \sum_{i=2}^{10} M_i \vec{v}_i$ of 9 planets (including Pluto) and the total angular momentum vector $\vec{L} = \sum_{i=1}^{10} M_i \vec{r}_i \wedge \vec{v}_i$ of the 10-body system [(Souami & Souchay 2012)].

The Z axis is directed by the direction of the total angular momentum $\vec{L}$. Two other vectors $\vec{C}$ and $\vec{D}$ that are related to $\vec{P}$ and $\vec{L}$ can now be defined by the following relations: $\vec{C} = \vec{P} \wedge \vec{L}$ and $\vec{D} = \vec{C} \wedge \vec{L}$. It is assumed that the X and Y axes are directed respectively along $\vec{D}$ and $\vec{C}$. The barycenter of the solar system, composed of 10 bodies in our calculation, is taken as the origin of our invariant reference system. As for the barycenter of the two-body Sun-Mercury system, it practically coincides with the center of the sun's position, and is located at a distance from the center of the sun of only about 1.4×10$^{-5}$ of the sun's diameter.

The maximum integration error of the MATLAB solver ODE113 has been previously analyzed [(Arminjon 2004), (Le Guyader 1993),(Roy 2014)], by comparing the difference between the initial values of the coordinates and velocities of the planets and their values after back and forth time integration. Figs. 1a and 1b show the relative error of the norm of the total angular momentum and the total mechanical energy with respect to the barycenter of the 10-body solar system, respectively. These quantities are conservative. It is clear that over a 700 year interval the magnitude of the total angular momentum and the total mechanical energy are respectively ~ 10$^{-14}$ and 10$^{-13}$ which shows the reliability of the calculation method.

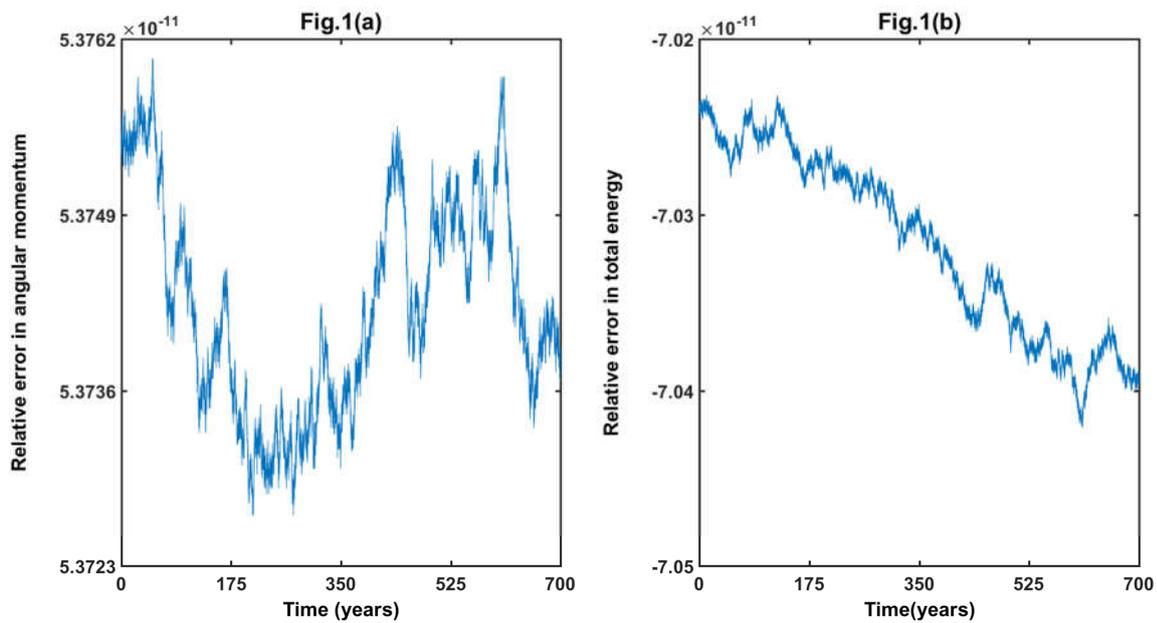

## 3. Results and Discussion

The position and velocity vectors uniquely describe the trajectory of the body in an inertial reference frame. The Newtonian gravitational equations allow a direct numerical integration of these quantities [(Aarseth 2003), (Everhart 1974), (Laskar 1988), (Tsiganis et al. 2005)]. In astrodynamics, it is common to use orbital elements instead of planetary positions and velocities which vary rapidly in space and time. The reason for this choice is the slow variation of the orbital elements and their simple geometrical interpretation, analogous to the Keplerian two-body problem, and assuming that the orbit for each instant is a conic section. The passage from the position-velocity space to the space of the orbital elements allows the solution of nonlinear differential equations by means of methods previously developed by Euler and Lagrange [(Taff 1985)]. The Keplerian orbital elements can be obtained from the position and velocity vectors by means of computer simulations [(Bate, Mueller & White 1971)]. However, the evolution of the solar system is strongly influenced by the interactions between the planets and therefore these orbital elements are time dependent, and therefore the perturbed motion is no longer Keplerian.

Reformulating the dynamic equations of planetary motion in terms of orbital elements results in an underdefined system, which means that additional constraints must be imposed in order to resolve the excess freedom [(Efroimsky 2002)]. Lagrange imposed the equality of the velocity vector of the real orbit and the velocity vector of the Keplerian orbit based on physical considerations and by the desire to simplify the calculations, known as Lagrange constraints

[(Taff 1985), (Chazy 1953), (Efroimsky & Goldreich 2004),(Gurfil 2004)]. The Lagrangian choice is the condition of the osculation. The dynamic behavior of the osculating orbital elements is then described by the Lagrangian planetary equations under the influence of a perturbing potential [(Brouwer & Clemence 1971), (Chazy 1953),(Taff 1985)]. It should be noted that there are other constraints that could be imposed to remove the extra degrees of freedom. As noticed by Efroimsky, the parameterization in terms of orbital elements is ambiguous [(Efroimsky 2002), (Battin 1999) p.477, (Brouwer & Clemence 1971) p. 273]. Due to a hidden symmetry in the Lagrange equations for N-body problems (N>2), the numerical calculation of the time evolution of the ellipses of variable shape followed by the orbiting bodies, can be a source of numerical instability [(Efroimsky 2002)]. In the description of the evolution of the orbital elements in time, the concept of osculating orbit is essential. The osculating orbit of a body in space at a given time is the Keplerian gravitational orbit, which would exist around the central body if the perturbations of all other bodies were absent. The osculating orbit (the instantaneous ellipse) is defined by the instantaneous heliocentric position vector and the instantaneous velocity at a point tangent to the physical trajectory. This means that the perturbed physical trajectory coincides with the Keplerian orbit that the body would follow if the perturbing force ceased instantaneously. The instantaneous heliocentric position vector and the instantaneous heliocentric velocity vector define a plane in space. There is infinity of ellipses passing through the same point and having the same velocity as the real orbit with foci that are not centered on the central body. As indicated by Efroimsky, the representation of the trajectory of the planet by a family of osculating ellipses is not unique [(Efroimsky 2002)]. Moreover, at least 5 points in space are needed to define an ellipse in a unique way.

Since the osculating orbit at each instant coincides with the orbit that the body would follow if the perturbations ceased instantaneously (elliptical orbit of the two-body problem), then in the N-body problem this can only be achieved if the velocities depend on the orbital elements in exactly the same way as in the two-body case. The description of the orbital evolution by osculating ellipses can then become even more complicated, because as will be shown in this paper, in the solar system, the sun is not exactly in one of the foci of a mean elliptical trajectory associated with the motion of each planet. Moreover, the real evolutionary orbit of Mercury is not elliptical but is a very complex three-dimensional trajectory, slowly evolving under the influence of the other planets of the solar system.

The evolution of the solar system is strongly modulated by the interactions between the planets. These interactions lead to an exchange of angular momentum and thus modify the orbital elements of the planets which then become time dependent. The inertial motion of the sun follows a complicated trefoliar-like quasi-symmetrical Jose cycle around the barycenter of the solar system [(Fairbridge & Shirley 1987),(Charvatova 1988)] and changes its position relative to the foci of the mean elliptical orbit of Mercury. The question is therefore how to extend the characteristics, such as foci and apsides of a Keplerian elliptic trajectory for an N-body problem. Strictly speaking, physical quantities such as eccentricity, major axis, and the foci of Keplerian elliptical trajectories are only defined for an ellipse, and therefore, for the actual orbits of the planets of the solar system, these physical quantities are invalid.

It should be noted that for Keplerian orbits, the apsides remain fixed in space and the line of junction of the apsides passes through the center of attraction. For the N-body problem, this is no longer valid. The line that joins the closest point to the sun to the farthest point from the sun no longer passes through the center of the sun. It rotates slowly by performing librations around the inertial reference frame associated with the barycenter of the solar system, thus breaking the axial symmetry of an elliptical orbit. Another interesting difference exists between the two-body and N-body Keplerian orbits concerning the closest and farthest points from the sun. For Keplerian orbits, the distance to the center of attraction is minimal and the velocity is maximal at the perihelion; conversely, the distance to the center of attraction is maximal and the velocity is minimal at the aphelion. For non-Keplerian orbits with N bodies, at the closest (or farthest) distance from the attracting body, the velocity of a planet is not maximal (or minimal). This is why; several characteristics of an open evolutionary orbit should be reconsidered, since their exact definition is only valid for a two-body Keplerian problem.

Different orbital characteristics of the two-body Keplerian problem will be now extended to the orbits of the 10-body solar system. For example, the sidereal orbital period of a planet can be defined as the time between two transits of the planet at the same point relative to the fixed stars. This definition must be slightly modified if the planet has an evolving orbit under the influence of other planets. The trajectory is open and therefore the planet passes through different points at each revolution. One can then introduce the concept of the mean Fourier period $T_{Fourier}$ as the dominant peak of the spectral power (Fourier transform) of the Mercury-Sun distance in a time interval allowing a very large number of revolutions of Mercury around the Sun. Note that during the time interval of 262144 days Mercury makes about 2980 revolutions around the Sun. It can be seen on Fig. 2a that the mean Fourier period as defined above is equal to $T_{Fourier}$=87.968 days.

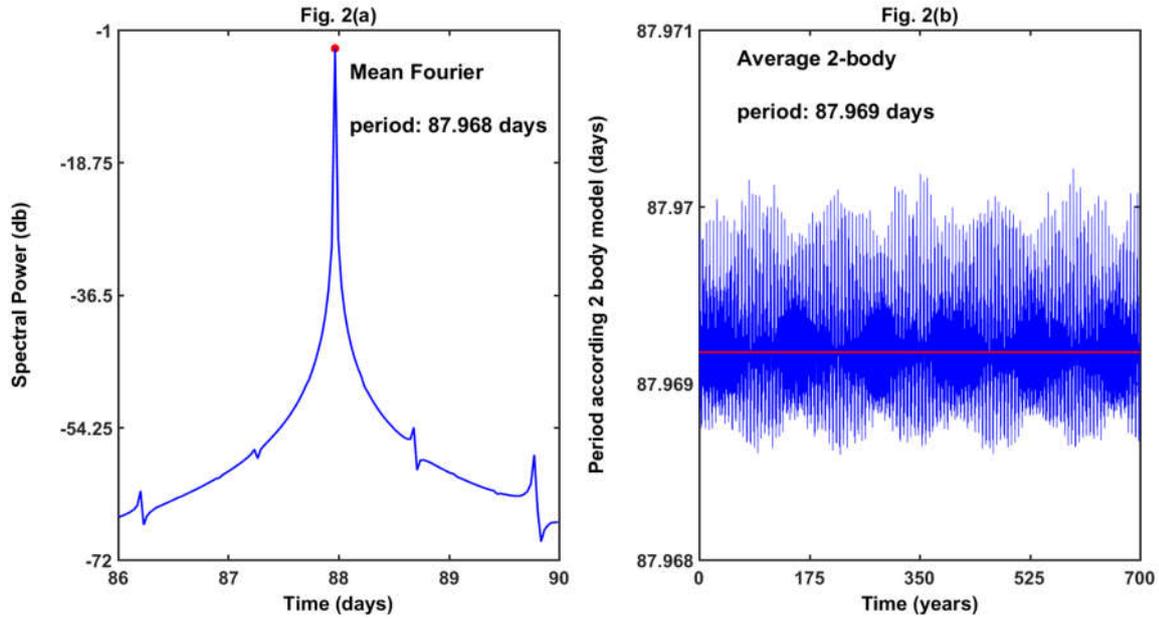

The chosen time interval $262144=2^{18}$ is convenient for the fast Fourier transform (fft). Note that this period is slightly longer than four Jose cycles: 262144 days ≈ 717.7 years ≈ 4×179 years.

Another mean orbital period $T_{2\text{-body}}$ can also be defined by analogy with the two-body problem as follows:

$$T_{2-corps} = \left(\frac{4\pi^2 a^3}{GM_\odot + GM_M}\right)^{1/2}. \quad (1)$$

where $M_S$ and $M_M$ are respectively the masses of the Sun and Mercury. G is the Newtonian constant of universal gravitation. For the two-body problem the semi-major axis $a$ is constant and can be defined at each point of the orbit as:

$$a = \frac{1}{\frac{2}{r} - \frac{v^2}{GM_\odot + GM_M}} \quad (2)$$

Here r is the distance and v the velocity of Mercury with respect to the sun in an arbitrary point of its trajectory. An analytical expression for two-body period as a function of r and v for each point of the trajectory can be obtained by inserting eq. (2) in eq. (1):

$$T_{2-corps} = \frac{2\pi(GM_\odot + GM_M)}{\left[\frac{2(GM_\odot + GM_M)}{r} - v^2\right]^{3/2}} \quad (3)$$

The averaged value of all these two-body periods in each trajectory point over the time interval 262144 days will be the average of two-body periods equal to : $T^m_{2-corps} = 87.969$ days as shown in Fig. 2b.

These two mean periods are slightly different, $T^m_{2-corps} \neq T_{Fourrier}$ and for a more exhaustive analysis, the larger of these two previous mean periods will be used, which allows us to always include a closest and a farthest point of approach to the sun in the interval of a mean period. It should be noted that depending on the number of iterations per year and the number of revolutions of Mercury taken into account, the maximum value is sometimes the mean Fourier period and sometimes the average two-body period. Once the mean orbital period has been chosen by the method described above, one can fix on for the initial reference time, an arbitrary instant which is different from the time of perihelion passage.

By definition, the perihelion of a planet in the two-body problem is the closest point of approach to the Sun of its elliptical trajectory. In such a definition of perihelion, all the points of the elliptical trajectory implicitly play a role since the perihelion must be the closest point to the sun among all the other points. The concept of perihelion can be extended to open trajectories close to the ellipse and known at least on one mean period (as defined above). Thus, the extended perihelion is the closest point to the sun among all the other points of a quasi-elliptical trajectory of the same mean period. In the same way, the extended aphelion is defined as the farthest point from the sun among all the other points of a quasi-elliptic trajectory of the same mean period. The direction of the extended major axis for each mean period will then be considered as the direction of the vector connecting the Sun to the extended perihelion. It should be noted that the lines joining respectively the extended perihelion and the extended aphelion to the Sun are not collinear. The average angle between these two directions is about 179.955°. Moreover the straight line joining the extended perihelion and the extended aphelion (extended line of apsides, or extended apside line) no longer passes through the center of the Sun. The extended perihelion and the extended aphelion have no fixed position and undergo slow and variable oscillations of very long periods [(Thornton & Marion 2004), page 312].

As can be seen in Fig. 3a (blue curve), the trajectory of Mercury even if it is close to an ellipse, it remains an open curve. The Z axis of Fig. 3a is drawn to the scale of the diameter of the Sun: D = 0.00465047 AU in order to visualize the opening of the ellipse of about $10^{-4}$ AU. The trajectory of the sun follows the red line on Fig. 3a. It is important to define the mean elliptical trajectory associated with the quasi-elliptical orbit of Mercury as a mean ellipse over the mean period in analogy with the two-body problem (cyan curve). Since all the points of the trajectory are not coplanar, a plane is first fitted to the points of the trajectory over one mean period starting from the initial reference time, as can be seen on Fig. 3b (the plane is shown in light brown color). All the points of the trajectory are projected into this plane and then an ellipse (in cyan color on Fig. 3b) is fitted to all the projected points of the mean period in the sense of a least squares fit. The foci of such a mean ellipse are then determined (star points in black color). On Fig. 3a the line of apsides of this mean ellipse is represented by the green segment. The major axis of the mean ellipse is collinear with the apsides line.

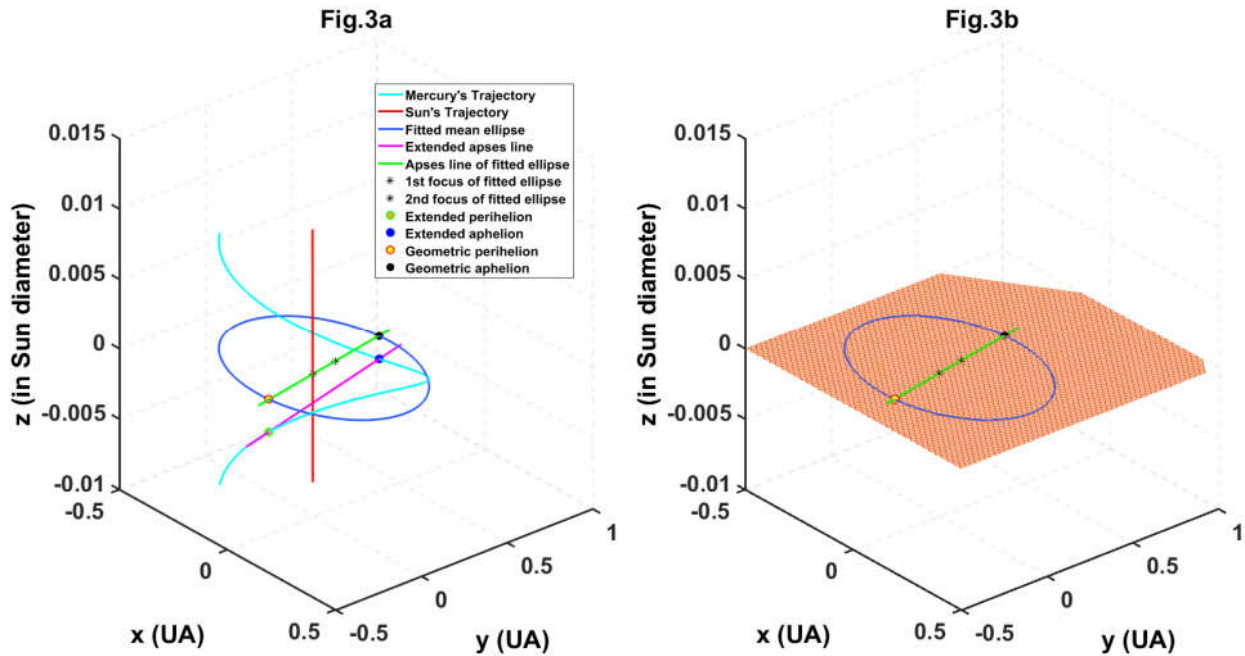

The extended line of the apsides joining the extended perihelion (orange point on the trajectory of Mercury Fig. 3b) and the extended aphelion (black point on the trajectory of Mercury) is represented by the magenta colored segment. It can be seen that the major axis of the mean ellipse and the extended line of the apsides are not parallel and are spatially separated in the scale shown on the Z-axis in Fig. 3a. Therefore, in addition to the extended perihelion, the geometrical perihelion will be defined as the point of the mean ellipse closest to the focus of the mean ellipse which is closest to the Sun. Geometrical perihelion is plotted on Fig. 3b, by a yellow color filled circle with red edge color, while the extended perihelion is represented on Fig 3a with orange color filled circle with green edge color. For two-body problems, the directions of the extended perihelion and the geometrical perihelion coincide. But for N-body problems, these directions are distinct, and their motion with respect to a inertial coordinate system must be evaluated separately.

It is important to underline that the notion of an average trajectory appears as soon as one deals with small deviations from a well identified behavior. For example, in the description of the slow evolution of the orbital elements, by the Lagrangian planetary equations, one uses the Lagrangian constraints. But instead of Lagrangian constraints, one can also use other constraints by reformulating new planetary equations called generalized gauge equations [(Efroimsky & Goldreich 2004)]. Gurfil et al[(Gurfil 2004)] extended the generalized constraints by presenting an averaged form of the gauge-generalized planetary equations with average classical orbital elements.

The advance of the extended perihelion and the geometrical perihelion can be characterized by the angle of rotation of the radius vectors pointing respectively to these perihelia with respect to a predefined reference direction. The advances of these extended and geometrical perihelia can be thus compared to the angle of rotation of the direction of the LRL vector with respect to this same reference vector.

In a large number of works [(Goldstein 1975), (Ebner 1985), (Davies 1983)] the advance of Mercury's perihelion is predominantly represented with the rotation of the LRL vector [(Goldstein, Poole & Safko 2008), (Goldstein 1975)]. In the two-body problem, this vector lies in the plane of the elliptical orbit, and is parallel to the major axis and points in the direction of perihelion [(Stewart 2005)]. Frequently, the rotation of the LRL vector is also used to estimate the contribution of general relativity to the perihelion precession of a planetary orbit [(Landau & Lifšic 2010), p370, (Weinberg 1972), p.230-233, (Garavaglia 1987), (Farina et al 2011)]. However, in the case of the N-body problem, the secular rotation of this vector cannot be clearly identified with the advance of Mercury's perihelion [(Ebner 1985)].

The LRL vector can be evaluated at each point of the trajectory as soon as the evolution of the position and velocity vectors in time is known. The LRL vector is a constant vector and is noted by $\vec{A}$ in this work. In the case of the two-body problem $\vec{A}$ is directed towards the perihelion and is parallel to the major axis :

$$\vec{A} = \frac{M_\odot M_M}{M_\odot + M_M} \left\{ \left( v_M^2 - \frac{GM_\odot + GM_M}{r_M} \right) \vec{r}_M - \left( \vec{r}_M \bullet \vec{v}_M \right) \vec{v}_M \right\} \quad (4)$$

where $\vec{r}_M$ and $\vec{v}_M$ are respectively heliocentric position and velocity vectors of Mercury.

In order to represent the rotation of this vector, it is necessary to follow its evolution with respect to a fixed reference direction. We can define the fixed direction of reference as the line of intersection of two instantaneous orbital planes: the initial osculating orbit plane of

Mercury and the initial ecliptic plane. The mean advance of the perihelion is generally estimated over a period of a century, during which Mercury makes 415 revolutions around the Sun. The linear least squares regression of the perihelion advance over a selected fitting time interval that may be longer than one century produces a trend line. Then the advance of the perihelion over a century is calculated by the help of linear fit coefficients. The angle of rotation between the direction of the LRL vector and the reference direction is shown in Fig. 4a. Fig. 4b shows a close-up picture of this evolution over a time interval of one century. This characteristic behavior is identical to the curves presented by Vankov [(Vankov 2010)] and Narlikar et al. [(Narlikar & Rana 1985)].

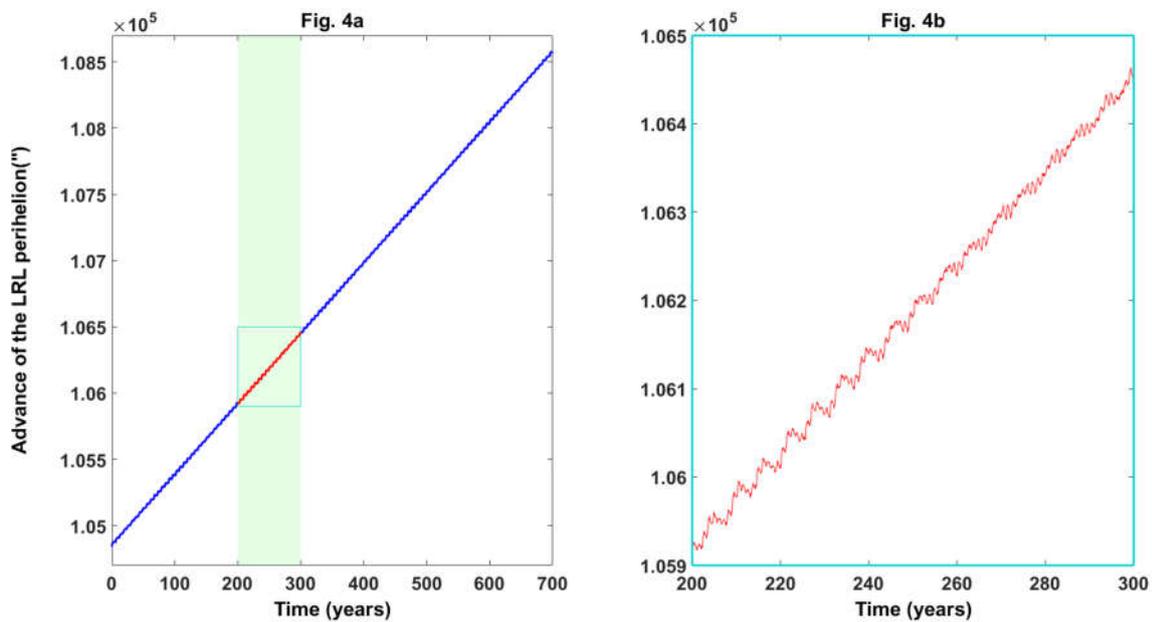

The advance of the perihelion as defined by the LRL vector will be compared with the advances of the extended and geometrical perihelia. For two-body problems, the direction of the LRL vector defined by Eq. (4), and the direction of the extended perihelia coincide.

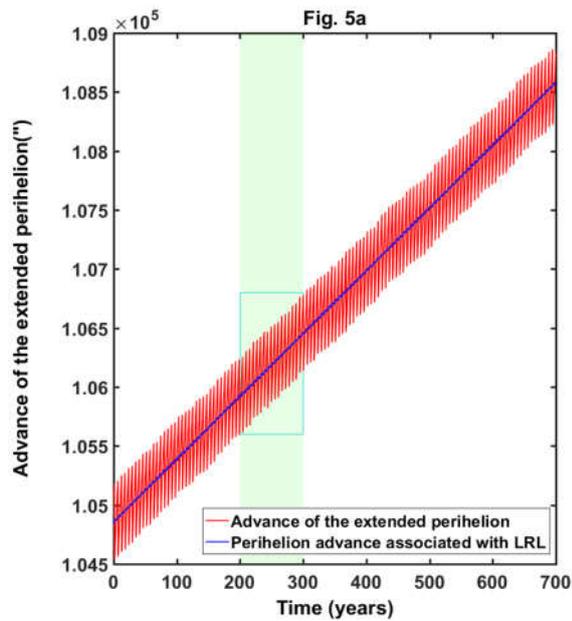
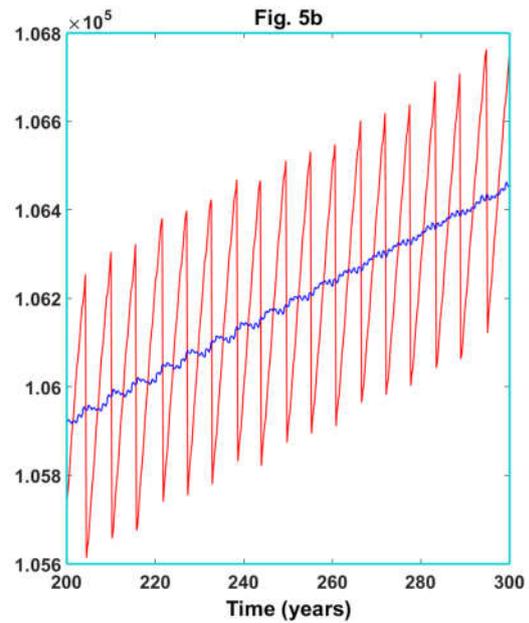

The advance of the extended perihelion can be calculated as the angle between the fixed reference direction and the position vector pointing to the extended perihelion at successive mean periods.

In Figs. 5a and 5b, an oscillatory behavior of the extended perihelion advance with an amplitude of about 700" can be observed. Also, a linear increase trend of the average advance of the extended perihelion can be noticed. The evolution of the LRL vector is added in blue in Figs. 5a and 5b for comparison purposes.

In addition to the extended perihelion, the geometrical perihelion is defined as the point of the mean ellipse closest to one of the foci (closest to the sun) of the mean ellipse. For the two-body problem, the direction of the LRL vector and the direction of the geometrical perihelion coincide.

The geometrical perihelion advance is calculated as the angle between the reference direction and the vector pointing from the focus closest to the sun of the mean ellipse to the point of the ellipse closest to this focus for each successive mean period. In Figs. 6a and 6b, the evolution of the advance of the geometrical perihelion is presented.

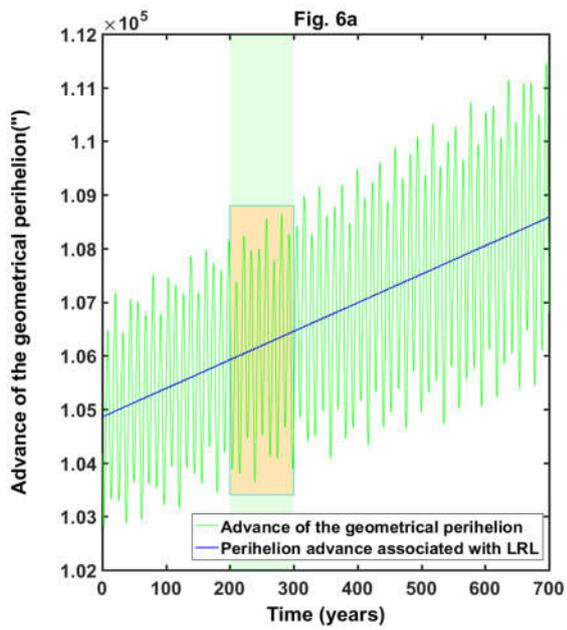
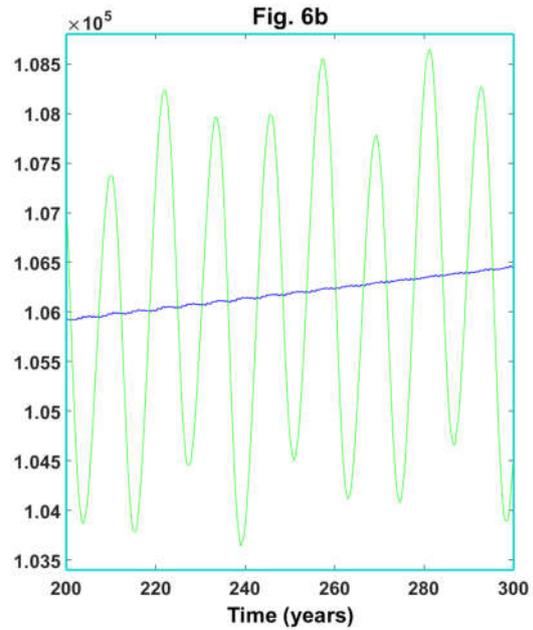

The mean amplitude of the oscillatory behavior of the geometrical perihelion advance is about 3700". One can observe a linear increase trend of the average geometrical perihelion advance. The time dependent LRL vector is also plotted in blue on Figs. 6a and 6b for comparison purposes.

In Figs. 7a, 7b and 7c, a Fourier analysis has been performed evaluating the spectral power of the perihelion advance associated with the rotation of the LRL vector with respect to the fixed reference direction as well as to extended and geometrical perihelion advances.

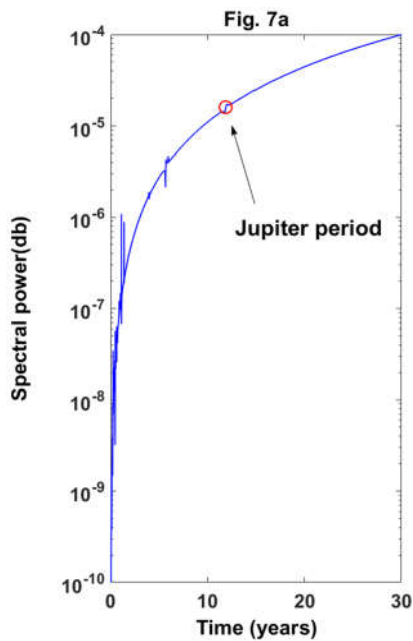
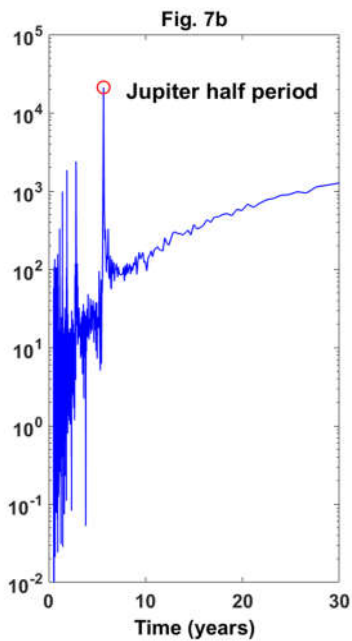
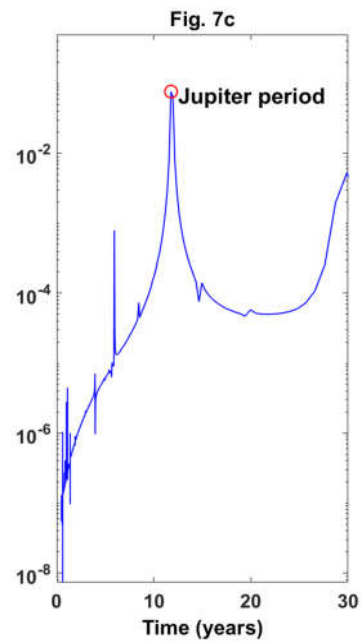

As can be seen on Fig. 7a the spectral power of the perihelion advance as defined by the LRL vector is almost independent from the motion of Jupiter while the extended and geometrical perihelion advances depend strongly on it (Fig. 7b and Fig. 7c). There is a striking difference between the spectral powers of the advances of the extended and geometrical perihelia, since the spectral power of the extended perihelion depends strongly on the half-period of Jupiter's rotation around the Sun (Fig. 7b), whereas that of the geometrical perihelion depends on Jupiter's period (Fig. 7c). It should be noted that the amplitude of their oscillatory time evolution behavior are different by a factor of about 5.

In different studies, the definition of the average perihelion advance per century is not clear. Of course, it is always the average value of the perihelion advance per century, but very often the authors have not provided details about the time interval of the evaluation of the average. Therefore, it is necessary to compare the influence of the fitting time interval on the mean perihelion advances per century associated with the rotation of the LRL vector, the extended and geometrical perihelion. The mean amplitude of the oscillatory behavior of the geometrical perihelion advance of Mercury is very large about 3700". The mean amplitude of the oscillations of the extended perihelion is about 700" (about 5 times smaller) while the mean amplitude of the oscillations of the advance of the perihelion associated with the LRL vector is only about 30". As mentioned before, the perihelion advance is fitted by the method of least squares to a linear function over a fitting time interval, in order to allow a description of the linear increase trend. Then the mean perihelion advance per century is deduced by coefficients of fitted liner function. If a very short adjustment interval (on the order of 100 years) is used for the calculation of the mean advance of the perihelion per century, the oscillatory behavior of large amplitude of the extended and geometrical perihelion advance will cause fluctuations of a very large amplitude, sometimes on the same order as the mean advance itself.

In order to describe the dependence of the mean perihelion advance per century on the fitting time interval, the evolution of the mean advance per century of the extended, geometrical and LRL vector defined perihelia for three different calculation intervals: 520 years, 718 years and 1011 years are respectively plotted in Figs. 8,9 and 10. For each interval, the calculation of the mean advance of the perihelion per century is started on an initial fitting period of 300 years which is gradually increased with a step of 5 years.

On Fig. 8a, 9a and 10a, the mean perihelion advances per century as calculated respectively for the LRL vector, for the extended perihelion and for the geometrical perihelion by choosing an integration period of 520years ($\approx 179\times 3$ years), with a time step of about 30 minutes have been plotted. In Figs. 8b, 9b and 10b the integration period is 718 years(≈179×4 years), with a time step of about 42 minutes. In Figs. 8c, 9c and 10c the integration period is 1011 years (≈179×6 years), with a time step of about 60 minutes.

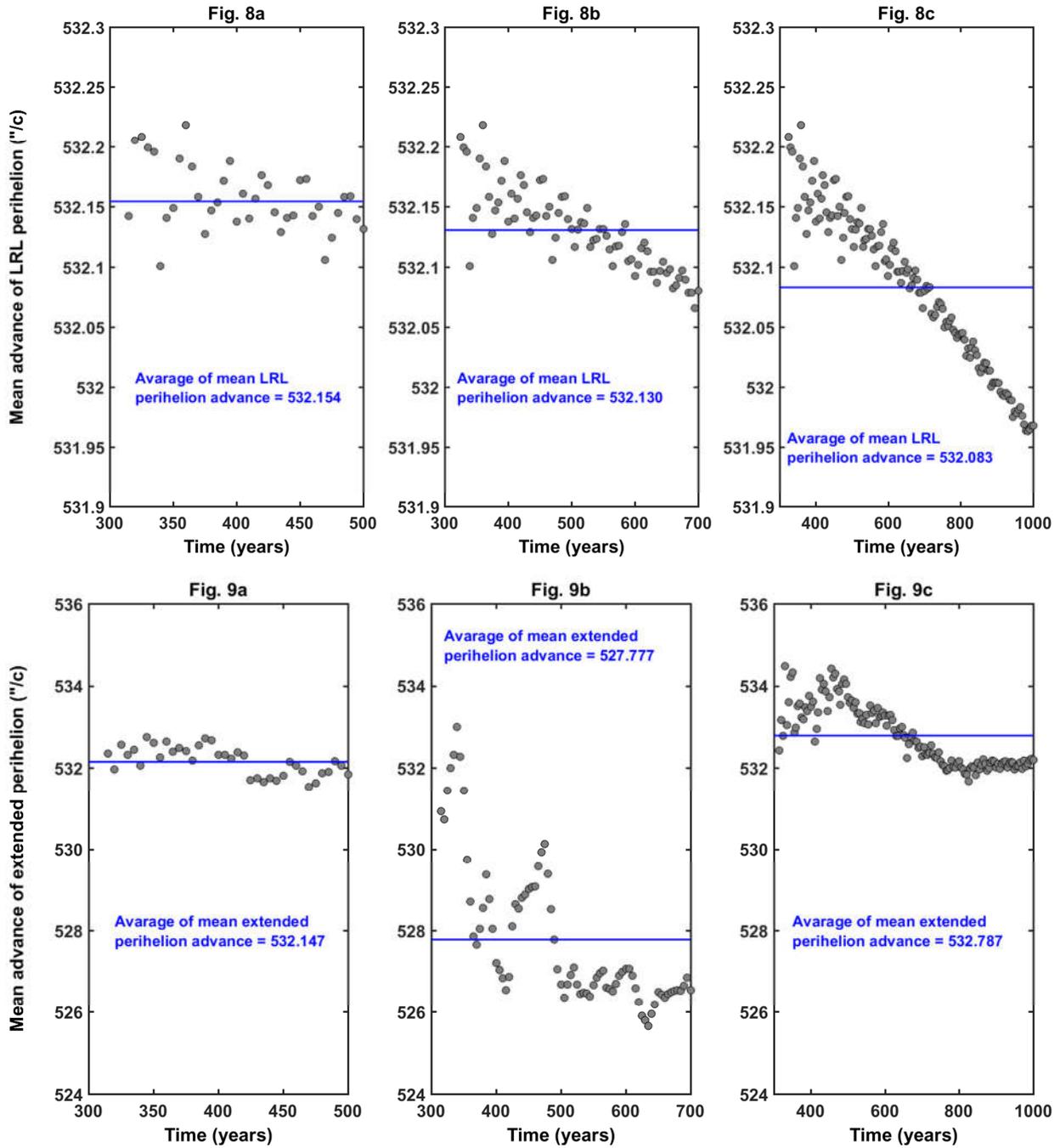

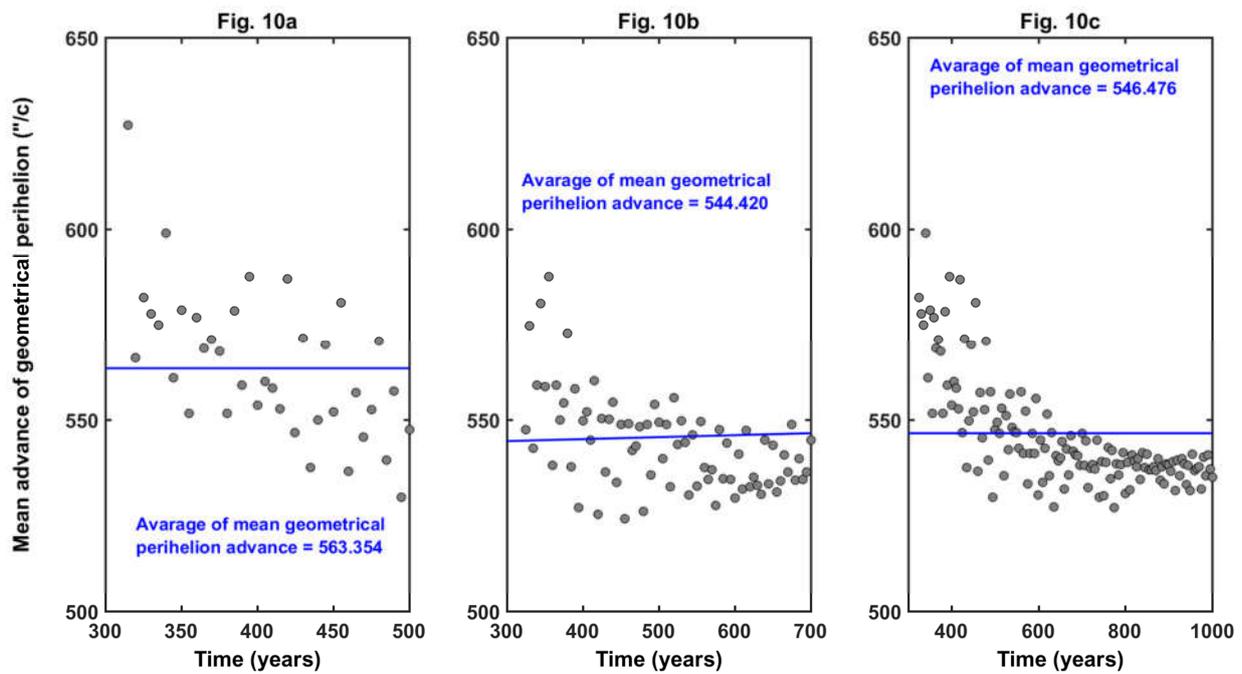

All the curves on Figs. 8 to 10 show a convergence tendency of the all mean perihelia advances per century towards a value of 532.1 "/c. It should be noted, however, that the mean perihelion advance per century as defined by the rotation of the LRL vector converges very rapidly to an average value of 532.1 "/c, while the convergence of the mean extended perihelion advance per century is slower. Naturally, the slowest convergence is that of the mean geometrical perihelion advance per century because of the oscillatory character of greater amplitude of the geometrical perihelion advance. It should be noted that the shape of the different mean perihelion advances per century for short fitting periods may depend on the number of points used in the integration. This dependence on the number of points used in the integration over an integration period of 1011 years, are illustrated on Figs. 11a, 11b and 11c, that have been calculated respectively with three different iteration steps : 12, 18 and 24 iterations per day. The mean perihelion advance per century as defined by the rotation of the LRL vector shows a very weak dependence on the time iteration step. The mean geometrical perihelion advance per century also shows also a very weak sensitivity to the time step. The mean extended perihelion advance per century is much more sensitive to the time step and for higher fitting time intervals there will be a convergence to the same value of 532.1"/c which was predicted by the rotation of the LRL vector.

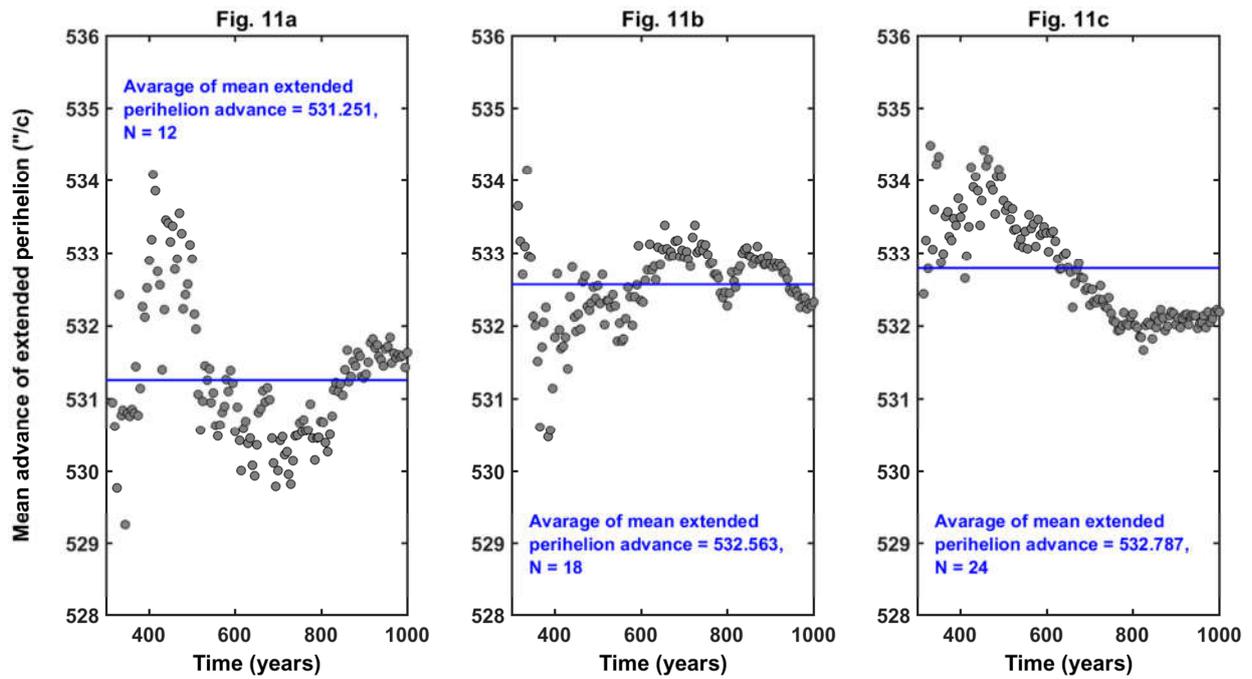

In perspective it is interesting to consider the long term behavior of mean perihelion advance per century, on a scale of time interval for a few million years. Future work will be needed to address this issue.

## 4. Conclusion:

The study of the mean perihelion advance of Mercury by means of numerical methods within the framework of a model including only the non-relativistic Newtonian gravitational interactions of the solar system: 9 planets (including Pluto) in translation around the Sun shows the importance of the exact definition of the latter because of the open and not elliptic trajectory of Mercury.

The notions of extended and geometrical perihelia have been introduced to distinguish more precisely the differences with the concept of the perihelion for the two-body problem and N body problem. Since the period of revolution of Mercury around the sun is no longer a strictly periodic function with a fixed period, the concept of the mean period of revolution has been introduced from Fourier analyses.

By fitting a mean plane to the trajectory of Mercury in a time interval of one mean period, a mean elliptical trajectory has been introduced whose focus and perihelion positions define the concept of a geometrical perihelion. The extended perihelion is defined as the closest point to the Sun each time interval of one mean period. The behavior of the advance of these new perihelia is compared with the rotation of the LRL vector.

I have shown in this study that the mean perihelion advance of Mercury per century as deduced from the behavior of the LRL vector, as well as the mean advances of the extended and geometrical perihelion per century, will depend on the fitting time interval and for intervals on the order of 1000 years, will converge to a value of 532.1"/c. Giant planets such as Jupiter can strongly influence the advance of extended and geometrical perihelia, but have little direct influence on the perihelion advance as calculated on the basis of the LRL vector rotation.

It was also shown that the mean advance of the extended perihelion per century depends strongly on the time step used in the calculations whereas the mean advance of the geometrical perihelion per century, as well as that deduced by the rotation of the LRL vector, will hardly depend on the time step used in our calculations.

**Figure Captions**

**Fig. 1.**

On Figs. 1a, and 1b, the relative error of the angular momentum and the mechanical total energy in the barycentric referential of the 10 body solar system are respectively plotted.

**Fig. 2.**

Fig. 2a shows the spectral power (in db) of the Fourier transform of the Mercury-Sun distance in a time interval of 262144 days which permits a very large number of revolutions (2980) of Mercury around the Sun. The red circle shows the position of the dominant peak. One can see on Fig. 2a, that the period defined by the dominant peak of the Fourier transform is equal to $T_{Fourier}$=87.968 days. The periods calculated from Eq. (3) based on the two-body model are presented in Fig. 2b. The average period obtained with this method gives: $T_{2-body}$=87.969.

**Fig. 3.**

On Fig. 3a the open trajectory of Mercury (curve in blue) in the time interval of one mean period is plotted. The line of the apsides of the fitted mean ellipse is represented by the green color segment. The extended line of the apsides joining the extended perihelion (orange point) to the extended aphelion (blue point) is represented by the magenta color segment.

In Fig. 3b, the mean orbital plane fitted to the points of the trajectory (which are not coplanar) in the time interval of one mean period is plotted in light brown color. An ellipse (shown in cyan on Fig. 3b) is fitted to all the trajectory points projected onto this mean plane in the sense of a least squares fit. The perihelion (white point) and aphelion (blue point) of this ellipse as well as its two foci (star points in black color) are also represented.

**Fig. 4.**

On Fig. 4a, the angle (in angular seconds) between the reference vector and the direction of the LRL vector is plotted as a function of time. In Fig. 4b, a close-up picture of this evolution in a time interval of one century is presented.

**Fig. 5.**

On Figs. 5a and 5b, the extended perihelion advance is plotted as a function of time. A close-up of the behavior of this evolution in a time interval of one century is illustrated on Fig. 5b. The time evolution of the LRL vector defined perihelion advance is also added in blue color on Figs. 5a and 5b for comparison purposes.

**Fig. 6.**

Figs. 6a and 6b demonstrate the behavior of the geometrical perihelion advance as a function of time. A closer look at the behavior of this evolution in a time interval of one century is illustrated on Fig. 6b. The time evolution of the LRL-vector-defined perihelion advance is also added in blue color on Figs. 6a and 6b for comparison purposes.

**Fig. 7.**

In Figs. 7a, 7b and 7c, a Fourier analysis of LRL, extended and geometrical perihelia are respectively performed. The spectral power (in db) of the advance of extended and geometrical perihelia (Figs. 7b and 7c) as well as the perihelion advance associated with the rotation of the LRL vector (Fig. 7a) with respect to the fixed reference direction as a function of the inverse of the frequency is plotted.

**Fig. 8.**

The perihelion advance deduced by the rotation of the LRL vector is fitted by the least squares method to a linear function. In Figs. 8a, 8b and 8c, the mean perihelion advance over a century is deduced by the fitting coefficients as a function of the fitting time interval for three different calculation intervals: 520 years, 718 years and 1011 years respectively. For each interval of calculation the procedure of calculation of the mean advance of the perihelion per century is started on an initial fitting time period of 300 years, which is gradually increased with a step of 5 years.

**Fig. 9.**

The advance of the extended perihelion is fitted by the least squares method to a linear function. In Figs. 9a, 9b and 9c, the mean perihelion advance over a century is deduced by the help of fitting coefficients as a function of the fitting time interval for three different calculation intervals: 520 years, 718 years and 1011 years respectively. For each interval the calculation of the mean advance of the perihelion per century is started on an initial fitting time period of 300 years, which is gradually increased with steps of 5 year intervals.

**Fig. 10.**

The advance of the geometrical perihelion is fitted by the least squares method to a linear function. In Figs. 10a, 10b and 10c, the mean perihelion advance over a century is deduced by the fitting coefficients as a function of the fitting time interval for three different calculation intervals: 520 years, 718 years and 1011 years respectively. For each interval, the calculation of the mean advance of the perihelion per century is started on an initial fitting period of 300 years, that is gradually increased it with a step of 5 years.

**Fig. 11.**

The dependence of the mean extended perihelion advance per century on the number of points used during the integration is illustrated on Fig. 11. For the integration period of 1011 years, three curves corresponding to three different iteration steps (12, 18 and 24 iterations per day) are respectively plotted in Figs. 11a, 11b and 11c.